\documentstyle[epsfig]{aa}

\newcommand{\be}{\begin{equation}}
\newcommand{\ee}{\end{equation}}

\begin{document}

\title{The afterglow of GRB021211: Another case of reverse shock emission}
\author{D.M. Wei\inst{1,2}}
\institute{Purple Mountain Observatory, Chinese Academy of Sciences,
Nanjing, China \and
   National Astronomical Observatories, Chinese Academy of Sciences,
   China}
\date{Received date/ Accepted date}

\abstract{GRB021211 was first detected by HETE II and its early
afterglow has been observed. There is a break in its afterglow light
curve at about 12 minutes after the bursts, before the break the
optical flux decays with a power-law index of about $-$1.6, while at
late time the power-law slope is about $-$1 (Chornock et al. 2002).
Here we will show that the afterglow light curve of GRB021211 can be
explained within the framework of the standard fireball model. We show
that the afterglow emission before the break time is the contribution
of the emission from both the reverse shock and the forward shock,
while the afterglow emission after the break time is mainly due to the
forward shock emission. From the fitting we can give constraints on the
parameters: the initial Lorentz factor $250\leq \gamma_0 \leq 900$, and
the surrounding medium density $n\geq 1.6\times 10^{-3}$ atoms ${\rm
cm^{-3}}$. We propose that since the values of $\epsilon_{\rm B}$ and
$\epsilon_{\rm e}$ are somewhat smaller for GRB021211, so the peak
energy of the reverse shock emission is well below the optical band,
and thus it is substantially fainter than 990123 at similar epochs.
Also we suggest that such a break might be a common feature in early
optical afterglows. } \maketitle

\section{Introduction}

GRB021211 is a bright, long gamma-ray burst detected by HETE II on
2002 December 11 at 11:18:34 UT. The burst duration in the 8 - 40
keV band was $>5.7$ seconds, the fluence was about $1 \times
10^{-6}$ ergs cm$^{-2}$ and the peak flux was $>8 \times 10^{-7}$
ergs cm$^{-2}$ s$^{-1}$ (Crew et al. 2002). This burst was also
observed by Ulysses and Konus - Wind. As observed by Ulysses, it
had a duration of about 15 seconds, a 25 - 100 keV fluence was
approximately  $1.8\times 10^{-6}$ erg cm$^{-2}$, and a peak flux
was about $4.5\times 10^{-7}$ erg cm$^{-2}$ s$^{-1}$ (Hurley et
al. 2002). The spectroscopic observations of the optical afterglow
identified three emission lines as [OII] 3727, and [OIII] 4959,
5007 at a redshift of $z=0.800 \pm 0.001$ (Vreeswijk et al. 2002).
Assuming $\Omega_{\rm m}=0.3$, $\Omega_{\rm \Lambda}=0.7$, and
$h=0.65$, the isotropic gamma-ray energy is $\sim 3.2\times
10^{51}$ ergs.

The prompt localization of GRB021211 by HETE II allowed the rapid
follow-up observation of the afterglow at very early time. Several
groups had detected the optical emission shortly after the gamma-ray
burst (Park et al. 2002; Li et al. 2002; Wozniak et al. 2002). The
observations show that the optical flux declined steeply at early time,
with a power-law index of about $-$1.6, while at later time the flux
decayed with a slope of about $-$1, the break time is about 12 minutes
after the burst (Chornock et al. 2002).

This break from a steep initial decline to a shallow later decline is
similar to the early behavior of GRB990123 (Akerlof et al. 1999). The
early emission of GRB990123 is believed to be due to the reverse shock,
while the later emission is ascribed to the normal forward shock (Sari
\& Piran 1999a). Recently Kobayashi \& Zhang (2002) have shown that the
re-brightening in the GRB021004 optical afterglow light curve can be
explained by the reverse shock and forward shock. Here we will show
that this early break around 12 minutes after the burst can be
interpreted as the superposition of the reverse shock and forward shock
emission.

\section{The emission from forward shock and reverse shock}

\subsection{Forward shock}

Multiwavelength follow-up of gamma-ray burst afterglows has
revolutionized GRB astronomy in recent years, yielding a wealth of
information about the nature of GRBs. The observed properties of GRB
afterglows are broadly consistent with models based on relativistic
blast waves at cosmological distances (Meszaros \& Rees 1997; Wijers et
al. 1997). In the standard fireball models, the huge energy released by
an explosion ($\sim 10^{52}$ ergs) is converted into kinetic energy of
a shell expanding at ultra-relativistic speed. After the main GRB event
occurred, the fireball continues to propagate into the surrounding gas,
driving an ultra-relativistic blast wave (forward shock) into the
ambient medium. The forward shock continuously heats fresh gas and
accelerates relativistic electrons to very high energy, which produce
the afterglow emission through synchrotron radiation.

Sari et al. (1998) have discussed the emission features of forward
shock in great details. Using their results, we have \be \nu_{\rm m, f}
= 5.1\times 10^{15}\left(1+z\right)^{1/2}\epsilon_{\rm
B}^{1/2}\epsilon_{\rm e}^2 g^2 E_{52}^{1/2}t_{\rm d}^{-3/2} \;\;\; {\rm
Hz} \ee \be \nu_{\rm c, f} = 2.7\times
10^{12}\left(1+z\right)^{-1/2}\epsilon_{\rm
B}^{-3/2}E_{52}^{-1/2}n^{-1}t_{\rm d}^{-1/2} \;\;\; {\rm Hz} \ee \be
t_{\rm m, f} = 2.9\left(1+z\right)^{1/3}\epsilon_{\rm
B}^{1/3}\epsilon_{\rm e}^{4/3} g^{4/3} E_{52}^{1/3}\nu_{\rm R,
15}^{-2/3} \;\;\; {\rm days} \ee \be t_{\rm c, f} = 7.3\times
10^{-6}\left(1+z\right)^{-1}\epsilon_{\rm B}^{-3}
E_{52}^{-1}n^{-2}\nu_{\rm R, 15}^{-2} \;\;\; {\rm days} \ee \be F_{\rm
\nu, max, f} = 110\left(1+z\right)\epsilon_{\rm
B}^{1/2}E_{52}n^{1/2}D_{28}^{-2} \;\;\; {\rm mJy} \ee where $\nu_{\rm
m, f}$ is the typical synchrotron frequency of forward shock emission,
$\nu_{\rm c, f}$ is the cooling frequency, $\epsilon_{\rm B}$ and
$\epsilon_{\rm e}$ are the fractions of the shock energy transferred to
the magnetic field and electrons, $g=(p-2)/(p-1)$, $p$ is the index of
electron energy distribution, $E_{52}$ is the burst energy in units of
$10^{52}$ ergs, $t_{\rm d}$ is the observer's time in units of 1 day,
$n$ is the surrounding medium density in units of 1 atom cm$^{-3}$,
$\nu_{\rm R, 15}=\nu_{\rm R}/10^{15}{\rm Hz}$, $D_{28}$ is the
luminosity distance in units of $10^{28}$ cm, $t_{\rm m, f}$ ($t_{\rm
c, f}$) is the time when the frequency $\nu_{\rm m, f}$ ($\nu_{\rm c,
f}$) crosses the observed optical frequency, and $F_{\rm \nu, max, f}$
is the peak flux.

According to the fireball model, before the peak time $t_{\rm m, f}$,
the observed optical flux is expected to increase as $F_{\nu}\propto
t^{1/2}$, when the typical synchrotron frequency crosses the observed
optical band, the flux reaches the maximum flux $F_{\rm \nu, max, f}$,
and then when $t > t_{\rm m, f}$, the flux decays as $F_{\nu}\propto
t^{-3(p-1)/4}$.

\subsection{Reverse shock}

The emission of reverse shock have been discussed by Meszaros \& Rees
(1997) and Sari \& Piran (1999b). This shock heats up the shell's
matter and accelerates its electrons, then these electrons loss energy
through synchrotron radiation. The reverse shock and the forward shock
are separated by a contact discontinuity, across which the pressure is
equal, so the energy density in both shocked regions is the same,
therefore the total energy in both shocks is comparable.

The typical synchrotron frequency $\nu_{\rm m, r}$ and cooling
frequency $\nu_{\rm c, r}$ of the reverse shock at the time when it
crosses the shell can be easily calculated by comparing them to those
of the forward shock. Since at the shock crossing time ($t_{\rm A}$),
the reverse shock and the forward shock have the same Lorentz factor
and energy density (which suggests the magnetic fields in both shocked
regions are the same, if we assume that the magnetic equipartition
factor is same in both regions), so the cooling frequency of the
reverse shock $\nu_{\rm c, r}$ is the same as that of the forward
shock, \be \nu_{\rm c, r}(t_{\rm A}) \simeq \nu_{\rm c, f}(t_{\rm A})
\ee The typical synchrotron frequency is proportional to the electrons
random Lorentz factor squared and to the magnetic field and to the
Lorentz boost. The Lorentz boost and the magnetic field are the same
for both the reverse shock and forward shock, while the random Lorentz
factor of reverse shock is $\gamma_0/\gamma_{\rm A}$ compared to
$\gamma_{\rm A}$ of the forward shock, i.e. the effective temperature
of reverse shock is much lower than that of the forward shock (by a
factor $\gamma_{\rm A}^2/ \gamma_0$, where $\gamma_0$ is the initial
Lorentz factor and $\gamma_{\rm A}$ is the Lorentz factor at the
crossing time.), then the typical synchrotron frequency of reverse
shock at the crossing time is \begin{eqnarray} \nu_{\rm m, r}(t_{\rm
A})& \simeq & \frac{\gamma_0^2}{\gamma_{\rm A}^4} \nu_{\rm m, f}(t_{\rm
A})\\ & \simeq & 3.5\times 10^{15}\epsilon_{\rm
B}^{1/2}\left(\frac{\epsilon_{\rm e}}{0.1}\right)^2 g^2
\left(\frac{\gamma_0}{300}\right)^2 n^{1/2} \, {\rm Hz} \end{eqnarray}
The peak flux at the typical frequency is proportional to the number of
electrons, the magnetic field and the Lorentz boost. The magnetic
fields and the Lorentz boost are the same for both reverse and forward
shock, while at the crossing time the mass of the shell is larger by a
factor of $\gamma_{\rm A}^2/\gamma_0$ than that of the ambient medium
swept by the forward shock (Kobayashi \& Zhang 2002), so we have
\begin{eqnarray} F_{\rm \nu, max, r}(t_{\rm A})& \approx &
\frac{\gamma_{\rm A}^2}{\gamma_0}F_{\rm \nu, max, f} \\ & \approx &
110\left(1+z\right)\frac{\gamma_{\rm A}^2}{\gamma_0}\epsilon_{\rm
B}^{1/2}E_{52}n^{1/2}D_{28}^{-2} \; {\rm mJy}\end{eqnarray} After the
reverse shock has passed through the ejecta, the ejecta cools
adiabatically. Sari \& Piran (1999a) have shown that, for $t
> t_{\rm A}$, the particle Lorentz factor evolves as $\gamma_{\rm e}\propto
t^{-13/48}$, the emission frequency drops quickly with time
according to $\nu_{\rm e}\propto t^{-73/48}$, and the peak flux
falls like $F_{\nu_{\rm e}}\propto t^{-47/48}$. So if the optical
band $\nu_{\rm R}$ is below the typical synchrotron frequency
$\nu_{\rm m,r}$, then the flux decays as $F_{\nu}\propto
t^{-17/36}$, while when $\nu_{\rm R}>\nu_{\rm m,r}$, the flux
decreases as $F_{\nu}\propto t^{-(21+73p)/96}$.

\section{Fitting the afterglow of GRB021211}

Using the emission features of reverse shock and forward shock
described above, we can fit the optical light curve of GRB021211. Here
we take the values $z=0.8$, $E=3.2\times 10^{51}$ ergs, and $p=2.3$.

For the forward shock emission, the observed optical flux is \be
\frac{F_{\rm \nu,f}(t)}{F_{\rm \nu,max,f}}=\left\{\begin{array}{ll}
\left(\frac{t}{t_{\rm m,f}}\right)^{1/2} & \;\;{\rm for}
\;\;\;t< t_{\rm m,f} \\
\left(\frac{t}{t_{\rm m,f}}\right)^{-3(p-1)/4} & \;\;{\rm for}
\;\;\;t_{\rm m,f}< t < t_{\rm c,f}
\end{array} \right . \ee Using eqs.(3)(5) and (11) we can give the
afterglow light curve from the forward shock, as shown in Fig.1 by the
dashed line. From fitting the observed data we can obtain the relation
\be \epsilon_{\rm B}\left(\frac{\epsilon_{\rm
e}}{0.1}\right)^{8/5}n^{3/5} \sim 9.1\times 10^{-4} \ee In addition,
the observation implies that $t_{\rm m, f}$ should be less than 100
seconds (if $t_{\rm m, f}>100$ s, there will be a bump in the afterglow
light curve), and $t_{\rm c, f}$ should be larger than 1 day (otherwise
there will be a steepening of the light curve) , so from eqs.(3)(4) we
have \be \epsilon_{\rm B}^{1/4}\left(\frac{\epsilon_{\rm
e}}{0.1}\right) \leq 0.14 \ee \be \epsilon_{\rm B}n^{2/3} \leq 0.04 \ee

For the reverse shock emission, for $t>t_{\rm A}$, we have the
relations $\nu_{\rm m,r}(t)=\nu_{\rm m,r}(t_{\rm
A})\left(\frac{t}{t_{\rm A}}\right)^{-73/48}$, $F_{\rm
\nu,max,r}(t)=F_{\rm \nu,max,r}(t_{\rm A})\left(\frac{t}{t_{\rm
A}}\right)^{-47/48}$, then the observed flux can be written as
\begin{eqnarray} F_{\rm \nu,r}(t)&=&F_{\rm \nu,max,r}(t)\left
[\frac{\nu}{\nu_{\rm m,r}(t)}\right ]^{-(p-1)/2}
\\&=&F_{\rm \nu,max,r}(t_{\rm A})\left [\frac{\nu}{\nu_{\rm m,r}(t_{\rm
A})}\right ]^{-\frac{(p-1)}{2}}\left(\frac{t}{t_{\rm
A}}\right)^{-\frac{73p+21}{96}}
\end{eqnarray} Using eqs.(8)(10) and (16) we can give the
afterglow light curve from the reverse shock, as shown in Fig.1 by the
dotted line. From fitting we can obtain the relation \be
\left(\frac{\gamma_0}{300}\right)^{23/10}\left(\frac{\gamma_{\rm
A}}{\gamma_0}\right)^2 \epsilon_{\rm
B}^{33/40}\left(\frac{\epsilon_{\rm
e}}{0.1}\right)^{13/10}n^{33/40}t_{\rm A}^2 \sim 9.8 \ee

\begin{figure}
\epsfig{file=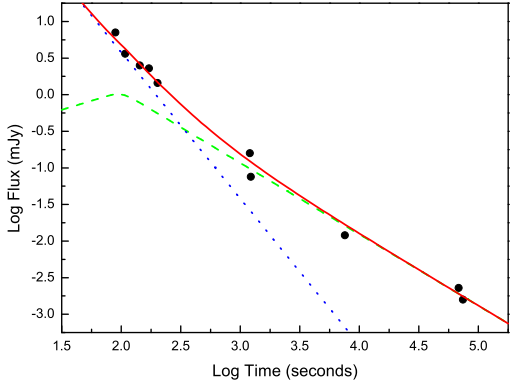, width=8cm} \caption{The optical light curve of
GRB021211. The dashed line is the emission of the forward shock, the
dotted line represents the emission from reverse shock, and the solid
line is the total flux. Data from: Price \& Fox 2002a, 2002b; Park et
al. 2002; Li et al. 2002; Kinugasa et al. 2002; McLeod et al. 2002;
Wozniak et al., 2002; Levan et al. 2002.}
\end{figure}

\noindent Combining eqs. (12)(14) and (17), we get \be
\left(\frac{\gamma_0}{300}\right)\left(\frac{\gamma_{\rm
A}}{\gamma_0}\right)^{20/23} \sim 31\epsilon_{\rm
B}^{-1/184}n^{-27/184}t_{\rm A}^{-20/23} \ee \be \epsilon_{\rm
B}^{1/16}\left(\frac{\epsilon_{\rm e}}{0.1}\right) \geq 0.077 \ee
Therefore eqs.(12)(13) and (19) give the constraint on the parameters
$\epsilon_{\rm B}$, $\epsilon_{\rm e}$ and $n$. Fig.2 shows the
relation between $\epsilon_{\rm B}$, $\epsilon_{\rm e}$ and $n$. The
dotted, dash-dotted, dashed and dot-dot-dashed lines represent $n=$0.1,
1, 10 and 0.0016 respectively. We find that the allowed values of
$\epsilon_{\rm B}$ and $\epsilon_{\rm e}$ lie in the region confined by
two lines Lc1 (eq.(19)) and Lc2 (eq.(13)). It is obvious that $n$ must
be larger than 0.0016, and $\epsilon_{\rm e}$ must be larger than
0.0077. If we take $n=1$, $\epsilon_{\rm e}=0.07$, then $\epsilon_{\rm
B}=1.6\times 10^{-3}$. We propose that more observations are needed in
order to further estimate the values of $\epsilon_{\rm B}$,
$\epsilon_{\rm e}$ and $n$.

\begin{figure}
\epsfig{file=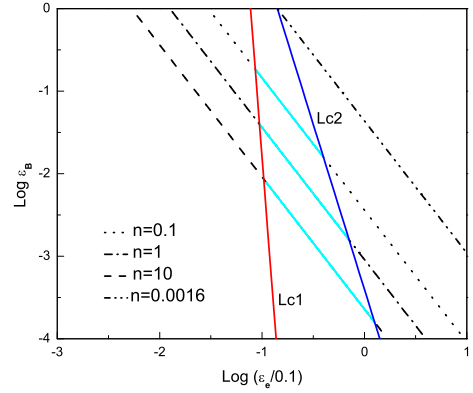, width=8cm} \caption{The relation between
$\epsilon_{\rm B}$, $\epsilon_{\rm e}$ and $n$ given by eqs. (12)(13)
and (19). The dotted, dash-dotted, dashed and dot-dot-dashed lines
represent $n=$0.1, 1, 10 and 0.0016 respectively. the allowed values of
$\epsilon_{\rm B}$ and $\epsilon_{\rm e}$ lie in the region confined by
two lines Lc1 (eq.(19)) and Lc2 (eq.(13)).}
\end{figure}

From eq.(18) we see that the initial Lorentz factor $\gamma_0$ depends
on $\epsilon_{\rm B}$ and $n$ very weakly, so as an approximation, and
taking $\gamma_{\rm A}\sim \gamma_0$, then we have $\gamma_0\sim
9300t_{\rm A}^{-20/23}$. Since the duration is about 15s and the first
observation time is 65s after the burst, so the value of $t_{\rm A}$
should lie between 15s and 65s, and therefore we can get the initial
Lorentz factor $250<\gamma_0 <900$, which is consistent with the lower
limit estimates base on the $\gamma$ - $\gamma$ attenuation calculation
(Fenimore et al. 1993).

\section{Discussion and conclusion}

The current afterglow observations usually detect radiation
several hours after the burst, at this stage the Lorentz factor is
independent of the initial Lorentz factor, thus these observations
do not provide useful information on the initial extreme
relativistic motion. The initial Lorentz factor is a very
important quantity for constraining the GRB models since it
specifies how "clean" the fireball is. Therefore to detect the
early afterglow of GRBs is very important, since it can provide
the information on the initial Lorentz factor. It is fortunately
that the early afterglow of GRB021211 have been observed, by
fitting its optical light curve we obtain its initial Lorentz
factor $250<\gamma_0 <900$, this value seems reasonable since it
is widely believed that the initial fireball Lorentz factor should
be larger than 100 in order to avoid photon-photon attenuation. Of
course, further constraint on the value of $\gamma_0$ needs more
early afterglow observations.

GRB990123 is the first burst for which its optical flash was observed,
the peak flux was about 1 Jy in R-band. After that many efforts have
been made to try to find the optical flash from other GRBs, but only
upper limits are given (Akerlof et al. 2000). Here we also note that
the optical flux of GRB021211 is substantially fainter than 990123 at
similar epochs. Why GRB990123 is so bright? One reason may be that
GRB990123 is a very bright burst, so its reverse shock emission is also
very strong. On the other hand, from fitting we note that the values of
$\epsilon_{\rm B}$ and $\epsilon_{\rm e}$ of GRB021211 are somewhat
smaller, which leads to the fact that the typical synchrotron frequency
of reverse shock is well below the optical band, so the early afterglow
(or optical flash) is weak. While for GRB990123 the typical synchrotron
frequency of reverse shock is close to the optical band (Sari \& Piran
1999a; Kobayashi \& Zhang 2002).

For smaller values of $\epsilon_{\rm B}$ and $\epsilon_{\rm e}$, not
only the typical synchrotron frequency of reverse shock is small, but
also the typical synchrotron frequency of forward shock is small, so
the time $t_{\rm m,f}$ when the typical frequency of forward shock
crosses the optical band is also small, for GRB021211, the observations
required $t_{\rm m,f}\leq 100$s. The late time afterglow for $t> t_{\rm
m,f}$ is $F_{\nu}=F_{\rm \nu, max, f}\left(t/t_{\rm
m,f}\right)^{-3(p-1)/4}$, so for smaller value of $t_{\rm m,f}$, the
observed optical flux should be much fainter than those with larger
values of $t_{\rm m,f}$, so we suggest that the so-called dark bursts
whose afterglow have not been observed might be due to their very small
values of $t_{\rm m,f}$.

The early afterglow of GRB021211 shows that there is a early break in
its optical light curve, before the break time the flux declined with a
power-law index of about $-$1.6, while at later time the flux decayed
with a slope of about $-$1. Although the reverse shock model predicts
that the optical flux should decay with a power-law index of about
$-$2, here we show that the superposition of both the forward shock and
the reverse shock emission can well account for the observed light
curve. Therefore we suggest that this early break might be a common
feature in early optical afterglow, and before the break time the slope
of flux decline may be flatter than $-$2 since it contains the
contribution from both the reverse shock and the forward shock
emission.

\acknowledgements  This work is supported by the National Natural
Science Foundation (grants 10073022 and 10225314) and the National 973
Project on Fundamental Researches of China (NKBRSF G19990754).


\begin{thebibliography}{}
\bibitem[]{} Akerlof, C., Balsano, R., Barthelemy, S., et al., 1999, Nature, 398, 400
\bibitem[]{} Akerlof, C., Balsano, R., Barthelemy, S., et al., 2000, ApJ, 532, L25
\bibitem[]{} Chornock, R., Li, W., Filippenko, A.V., \& Jha, S., 2002,
GCN 1754
\bibitem[]{} Crew, G., Villasenor,J., Vanderspek,R., et al., 2002, GCN 1734
\bibitem[]{} Fenimore, E.E., Epstein, R.I., \& Ho, C., 1993, A\&AS, 97,
59
\bibitem[]{} Hurley, K., Mazets, E., Golenetskii, S., \& Cline,
T., 2002, GCN 1755
\bibitem[]{} Kinugasa, K., Kato, T., Yamaoka, H., \&  Torii, K., 2002,
GCN 1749
\bibitem[]{} Kobayashi, S., \& Zhang, B., 2002, astro-ph/0210584
\bibitem[]{} Levan, A., Fruchter, A., Welch, D., et al., 2002, GCN 1758
\bibitem[]{} Li,W., Filippenko, A.V., Chornock,R., \& Jha, S., 2002,
GCN 1737
\bibitem[]{} McLeod, B., Caldwell,N., Grav, T., Luhman,K., Garnavich,P., \& Stanek, K.Z., 2002, GCN 1750
\bibitem[]{} Meszaros, P., \& Rees, M.J., 1997, ApJ, 476, 231
\bibitem[]{} Park,H.S., Williams,G.,\& Barthelmy, S., 2002, GCN 1736
\bibitem[]{} Price, P.A., \& Fox, D.W., 2002a, GCN 1732
\bibitem[]{} Price, P.A., \& Fox, D.W., 2002b, GCN 1733
\bibitem[]{} Sari, R., Piran, T., \& Narayan, R., 1998, ApJ, 497, L17
\bibitem[]{} Sari, R., \& Piran, T., 1999a, ApJ, 517, L109
\bibitem[]{} Sari, R., \& Piran, T., 1999b, ApJ, 520, 641
\bibitem[]{} Vreeswijk, P., Burud, I., Fruchter, A.,\&
Levan, A., 2002, GCN 1756
\bibitem[]{} Wijers, A.M.J., Rees, M.J., \& Meszaros, P., 1997, MNRAS,
288, L51
\bibitem[]{} Wozniak, P., Vestrand, W.T., Starr D., et al., 2002, GCN 1757
\end{thebibliography}
\end{document}